\documentclass[letter,11pt]{article}
\usepackage{jinstpub}
\usepackage{booktabs}
\usepackage{times}
\usepackage{subfig}
\usepackage{caption}
\usepackage{upgreek}
\usepackage[draft]{todonotes}

\title{High dynamic range CdTe mixed-mode pixel array detector (MM-PAD) for kilohertz imaging of hard x-rays}

\author[a]{Hugh T. Philipp}
\author[a]{Mark W. Tate}
\author[a]{Katherine S. Shanks}
\author[a]{Prafull Purohit}
\author[a,b]{Sol M. Gruner}

\affiliation[a]{Laboratory of Atomic and Solid State Physics\\Cornell University\\Ithaca, NY 14853, U.S.A.}

\affiliation[b]{Cornell High Energy Synchrotron Source (CHESS)\\Cornell University\\Ithaca, NY 14853, U.S.A.}

\emailAdd{htp2@cornell.edu}

\abstract{A hard x-ray, high-speed, high dynamic range scientific x-ray imager is described. The imager is based on the mixed-mode pixel array detector (MM-PAD) readout chip coupled to a 750\,$\upmu$m thick cadmium telluride (CdTe) sensor. The full imager is a $2\,\times\,3$ tiled array of MM-PAD sensor/readout chip hybrids. CdTe improves detection for high energy x-rays as compared to silicon sensors, enabling efficient x-ray imaging to extend to >100\,keV. The detector is capable of 1\,kHz imaging and in-pixel circuitry has been designed to allow for well depths of greater than $ 4 \times 10^{6}$ 80\,keV x-rays. A charge integrating front-end allows for quantitative measurement of high flux x-ray images beyond the capabilities of photon counting detectors. Detector performance is summarized and experimental measurements are presented.}


\begin{document}
\maketitle

\section{Introduction}

The mixed-mode pixel array detector (MM-PAD) is a wide dynamic range hybrid scientific x-ray imaging device \cite{Tate2013} that, when coupled to a silicon sensor, has demonstrated utility for scientific investigations at synchrotron sources~\cite{Giewekemeyer2014,Giewekemeyer2019,Ayyer2014,Liu2017,Overdeep2018}. The same silicon sensor hybrid device is also the basis of the electron microscope pixel array detector (EM-PAD)~\cite{Tate2016} that has recently acquired record breaking microscopy resolution~\cite{Jiang2018}. The MM-PAD readout chip is distinguished by in-pixel signal processing electronics that combine a charge integrating front-end with threshold triggered charge removal circuitry and a digital counter to extend the single-image well-depth while retaining low noise performance. 
This pixel architecture can record accurate flux measurements at high intensities that are beyond the capabilities of photon-counting detectors while maintaining single-photon signal resolution. In addition, the charge integrating architecture used here is not susceptible to systematic errors which can arise in photon-counting detectors when events have charge-splitting between pixels, such as under- or over-counting. This paper reports the specific capabilities of the Cornell-produced CdTe MM-PAD detectors that are in use at both the Advanced Photon Source (APS) and Cornell High Energy Synchrotorn Source (CHESS). 

\begin{figure}[t!]
    \centering
    \includegraphics[width=0.9\textwidth]{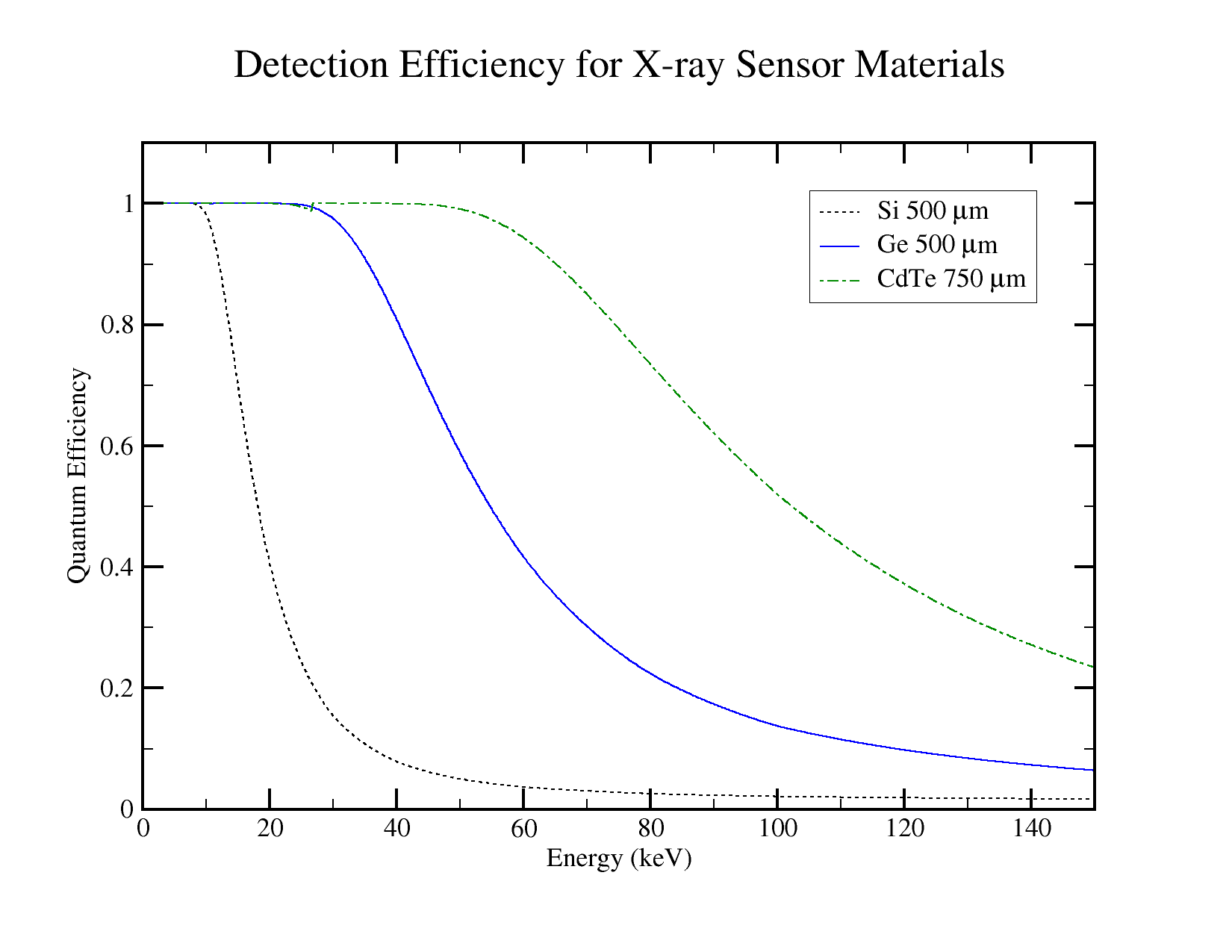}
    \caption{Efficiency curves for a 500\,$\upmu$m thick silicon sensor compared to a 750\,$\upmu$m thick CdTe sensor. These are the standard sensor thicknesses used with the MM-PAD chip. Efficiency for a 500\,$\upmu$m thick Ge sensor is also plotted for reference. At this point, no germanium sensor is available. }
    \label{fig:efficiency}
\end{figure}

Silicon has served as a high quality x-ray and electron detection layer for the MM-PAD and many other hybrid devices. The material properties of silicon are well suited to provide high fidelity direct detection of x-rays for energies below 20\,keV. At 20\,keV, where silicon has an absorption depth approaching 1\,mm, the practical usefulness of silicon is greatly diminished (a typical silicon sensor thickness is 500\,$\upmu$m). Cadmium telluride (CdTe), when compared to silicon, has the huge advantage of offering high-efficiency direct detection of high energy x-rays with an absorption depth of approximately 1\,mm at 100\,keV. A plot of efficiencies for selected materials and sensor thicknesses is shown in Figure~\ref{fig:efficiency}.This extended x-ray energy range is important for studies in material science where properties of physically large samples and/or high atomic number materials are investigated~\cite{Chatterjee2017,Chatterjee2019}. CdTe has been extensively studied for x-ray and gamma-ray detection. The main advantage, high energy detection efficiency, is a direct result of high atomic numbers of cadmium (Z = 48) and tellurium (Z=52). Additionally, a relatively large bandgap of 1.44 eV leads to low dark current compared to silicon (bandgap of 1.12\,eV) and makes it suited for charge integrating front-end electronics. At the same time, CdTe suffers from charge trapping in the bulk material which leads to material polarization. Charge trapping and material non-uniformity/crystal defects lead to variable lateral fields that make charge collection a function of x-ray interaction depth. Additionally, CdTe fluoresces with K$_{\upalpha}$ energies of 27.5\,keV for Te and 23.2 for Cd. The mean free paths of these fluorescent x-rays are approximately 62.6\,$\upmu$m and 118.8\,$\upmu$m respectively in CdTe, meaning that for pixel sizes on the order of 100\,$\upmu$m there is an expectation of non-ideal imaging effects caused by fluorescent x-ray photons exiting pixels after the initial x-ray interaction with the sensor. These material effects have been extensively explored and summarized in the literature~\cite{Limousin2003,Becker2016b,Becker2017,Aamir2011}. Despite the drawbacks of CdTe, it remains one of the best options for high-energy direct detection of x-rays and, as described in this paper, can be used to make a scientifically productive x-ray imager when combined with a high dynamic range readout chip. 

A summary of the MM-PAD capabilities are shown in Table \ref{Tab:specs}, where figures for both the silicon and CdTe sensor equipped detectors are offered for comparison.

\begin{table}[htbp]
\caption{Specifications MM-PAD with Si and CdTe sensors.}
\centering
\begin{tabular}{l l l}
\toprule
Specification & CdTe MM-PAD & Si MM-PAD \\
\cmidrule(r{4pt}){1-3}
Pixel Size  & \multicolumn{2}{l}{\hspace{18pt}both $150\, \upmu$m$\, \times\, 150\, \upmu$m} \\
Format (2\,$\times$\,3 tiling) & \multicolumn{2}{l}{\hspace{18pt}both 256 pixels$\, \times$ 384  pixels} \\
\cmidrule(r{4pt}){2-3}
Read Noise & 1.6 keV equivalent & 1.3 keV equivalent \\
Full Well (absorbed x-ray energy) & $ 4.6\, \times 10^{8}$ keV & $3.8\, \times 10^{8}$ keV \\
Maximum Sustained flux & $3.9\, \times 10^{9}$ keV/s & $3.2\, \times 10^{9}$ keV/s\\
Sensor Thickness & 750$\,\upmu$m & 500 $\,\upmu$m \\ 
\end{tabular}
\label{Tab:specs}
\end{table}

\section{Detector Description}

\begin{figure}[b!]
    \centering
    \includegraphics[width=0.55\textwidth]{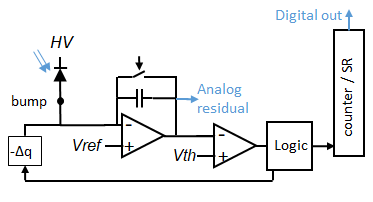}
    \caption{High-level schematic of the MM-PAD pixel. A charge integrating amplifier is monitored by a voltage comparator, which, when triggered, will increment a digital counter and initiate a charge removal event ( $\sim$ 1.6\,MeV equivalent charge) as the integration continues. At the end of each frame, each pixel yields 18 bits of digital data and an analog residual, digitized off-chip to 12 bits. These two numbers are scaled and combined to give a 30 bit intensity/pixel.}
    \label{fig:mmpix}
\end{figure}

The application specific integrated circuit (ASIC) is the same used in the silicon version of the MM-PAD\cite{Tate2013} and the EM-PAD\cite{Tate2016}. The ASIC is fabricated in the Taiwan Semiconductor (TSMC) 0.25\,$\upmu$m mixed-mode process with metal-insulator-metal (MiM) capacitors. The pixel pitch is 150\,$\upmu$m on a regular grid with each individual ASIC having 128\,$\times$\,128 pixels arranged in 8 banks of 16\,$\times$\,128 pixels. A high-level diagram of the pixel is shown in Figure~\ref{fig:mmpix}. A charge integrating amplifier accumulates the signal generated in the sensor diode onto a small (50 fF) capacitor. The small feedback capacitance is chosen to provide single photon sensitivity at low intensity levels. The output of the amplifier is monitored by a comparator during integration. When a preset threshold is exceeded during exposure, logic increments a counter and initiates an event which removes a fixed amount of charge, $\Delta$q, from the input node. This keeps the charge integrator within its operating range, greatly increasing the dynamic range of the pixel. Integration from the sensor continues during each charge removal event in a dead-timeless process. Response continues to be linear up to a maximum input rate of $3.9\, \times 10^{9}$ keV/pixel/s. Pixel output at the end of a frame is mixed mode, consisting of an analog value that is sampled from the front-end integrator and the 18-bit digital word that tracks the number of charge removal events \cite{SchuettePHD}. The analog value is digitized with a 12-bit analog-to-digital converter (ADC) located external to the chip, running at 2.5\,MHz. The x-ray equivalent noise of the digitized analog signal is 1.6\,keV when using a CdTe sensor. 
The 12-bit value from the analog residual and the 18-bit in-pixel counter are scaled and combined to yield a 30 bit intensity value per pixel per frame. Full calibration of the detector requires scaling the two parts of the 30-bit digital word together to yield a continuous, linear response.
The analog readout uses a row-select combined with a per-bank analog multiplexer to cycle through columns. The full ASIC is read in 860 microseconds, during which time the ASIC cannot acquire additional charge and the front-end of the detector is held in reset. The ASIC readout and the wire bonds that supply all power and necessary control signals to the chip are restricted to one edge of the ASIC, allowing the chip to be abutted on 3 sides when tiled in a larger array, minimizing inactive area. 

\begin{figure}[t!]
    \centering
	\includegraphics[width=0.8\textwidth]{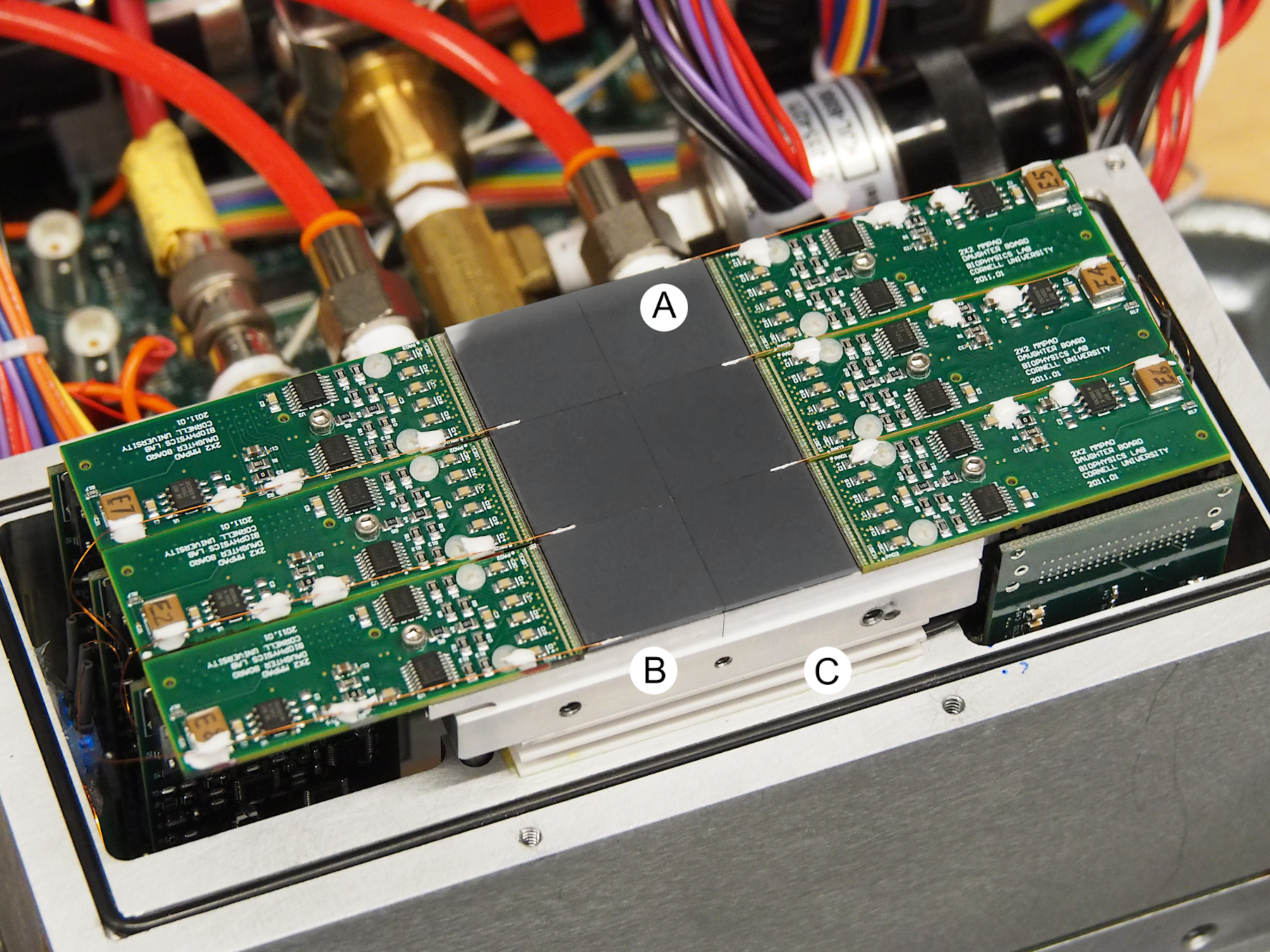}
    \caption{A 2\,$\times$\,3 tiling of MM-PAD CdTe modules in a vacuum housing with front face removed. Each module has a dedicated printed circuit board mounted to a secondary ADC board at right angles which in turn mates to a backplane board. A bias contact to the front of each sensor is made with a length of 38 gauge magnet wire. This wire is attached with a small amount of silver paint in an area of the sensor that extends beyond the imaging area. The magnet wire is attached to the associated circuit board with epoxy for mechanical stability of the assembly. The wire extends off the circuit board to a bias distribution point to provide the ability to bias the sensors up to 1000\,V. The CdTe hybridized modules, (A), are in thermal contact with an aluminum heatsink (B). The temperature of the heat sink is monitored and actively controlled with a thermoelectric cooler (C). Water is circulated through a portion of the detector housing to provide heat transfer.}
    \label{fig:mmpic}
\end{figure}


The fully tiled array, designed and used by the Cornell detector group, consists of 6 ASICs tiled in a 2\,$\times$\,3 array with each ASIC having a dedicated pixellated CdTe Schottky diode matching the 128\,$\times$\,128 pixels of the ASIC. The tiled detector has 256\,$\times$\,384 active pixels. In practice, the tiling of the diodes, associated guard rings, and assembly requires a small inactive gap of 5 to 6 pixels between sensor modules. The CdTe-indium Schottky hole-collecting diode is manufactured by Acrorad (Okinawa, Japan) and the bump-bonding of the CdTe to the ASIC is performed by Oy Ajat (now Direct Conversion, Espoo, Finland). Each ASIC in the tiled array has a dedicated aluminum heat sink and printed circuit board. The ASIC-CdTe hybrid is attached to the heat sink using a pattern of thermally conductive silicone (Nusil R-2930). The hybridized ASIC, associated printed circuit board, and the heat sink comprise the detector module which is complete after the ASIC is wire bonded and a high voltage bias line is provided for sensor biasing. Six modules are mounted on a larger aluminum heatsink block. For the CdTe system, the temperature is typically maintained at 0\,C with a thermoelectric cooler. The modules and the entire inside of the detector housing is kept at a vacuum to prevent the condensation of water and to thermally isolate the detector modules. Thin layers of Apiezion N vacuum compatible grease is used between heatsinks to improve thermal contact.  
\begin{figure}[t!]
    \centering
    \includegraphics[width=0.9\textwidth]{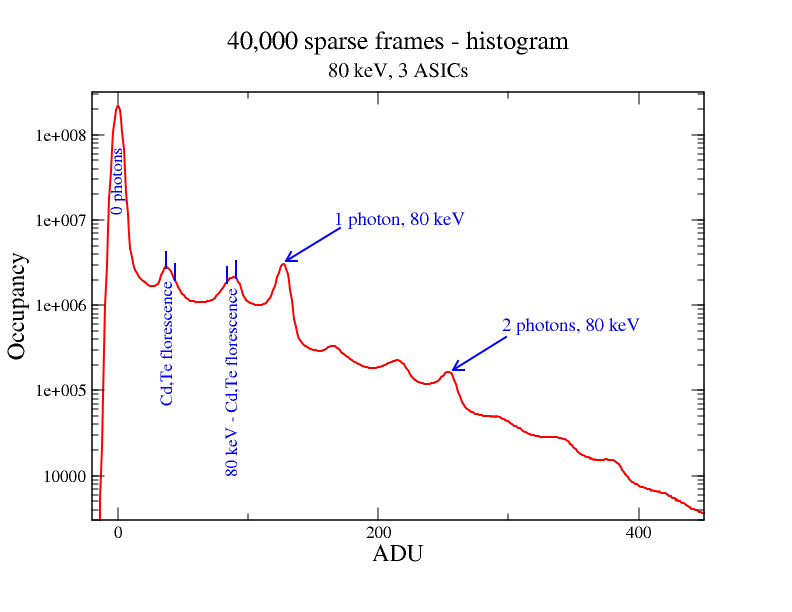}
    \caption{Histogram of detector output for a sparse data set over 40,000 frames. The x-ray energy is 80\,keV and was collected at APS 1-ID-E. Fluorescent and escape peaks, corresponding to 80\,keV minus the fluorescence energy, are labeled. Fluorescence for Cd and Te occur at 23.2\,keV and 27.5\,keV respectively.}
    \label{fig:mmhist}
\end{figure}

The modules are mated to six secondary circuit boards with 8 ADC channels/board. These in turn mate to a large backplane board that serves to route all electrical signals to and from the modules and also provides a vacuum feed through by being sandwiched between two sections of housing with o-rings providing the vacuum seal.

The full detector data stream is assembled and organized by a dedicated FPGA that rearranges bits and translates the pseudo-random in-pixel pixel counter outputs to integer values. Once the analog and digital information is properly arranged, the data is conveyed to a PC via a CameraLink connection. The maximum frame rate is 1.1\,kHz as dictated by the 860\,$\upmu$s readout. 

An assembled MM-PAD ASIC 2\,$\times$\,3 tiling with CdTe sensor is shown in Figure~\ref{fig:mmpic}.




\section{Performance}


This detector has been used in several high-energy x-ray experiments
at the APS and CHESS. One experiment\cite{Chatterjee2019}, collected 4 dimensional
data sets, tracking the dynamic behavior of individual grains in
a sample of titanium and magnesium allows under stress at the 1-ID-E beamline
at APS. Within these data sets, much of the area of the recorded diffraction in each
image had $<< 1$\,photon/pixel, allowing a spectral analysis of the recorded signal from the incident 80\,keV x-rays, shown as a logarithmic plot in Figure~\ref{fig:mmhist}. X-ray fluorescence peaks and associated fluorescent escape peaks are clearly visible with the corresponding cadmium and tellurium fluorescent energies and escape peaks marked. 





\begin{figure}[b!]
    \centering
    \includegraphics[width=0.9\textwidth]{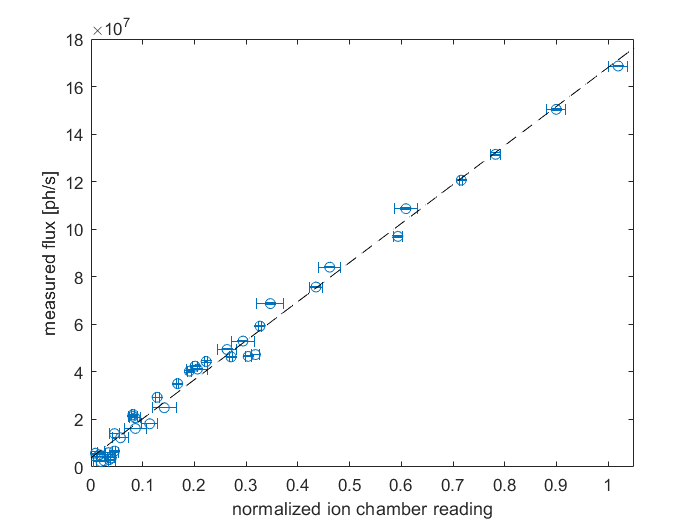}
    \caption{Measurement of integrated spot intensity on the CdTe MM-PAD as a function of normalized ion chamber output. }
    \label{fig:highflux}
\end{figure}

Performance of the detector at high flux was tested at APS beamline 6-ID-B. The detector was mounted in the path of the direct beam and the direct beam was imaged with no beamstop as a series of aluminum attenuators was used to vary the transmission from $9 \times 10^{-4}$ to $0.05$. A Si <111> monochromator was used to set the photon energy to 13 keV. Although this energy is within the range of a detector with a Si sensor, it is also well within the sensitivity range of CdTe and is useful to interrogate the high-flux behavior of this detector. The incident intensity was monitored by an ion chamber 325\,cm upstream of the detector. Slits immediately upstream of the ion chamber restricted the x-ray spot size to $100\,\upmu$m$\,\times\,800 \upmu$m. With beam divergence, the spot size on the detector face was approximately 330\,$\upmu$m$\,\times\,970\,\upmu$m FWHM. At each attenuator setting, 200 frames with 100\,ms integration time were acquired. The detector response was linear over the flux range measured, as shown in Figure \ref{fig:highflux}. Due to a lack of absolute ion chamber calibration, the ion chamber readings have been normalized to the average reading recorded at the maximum transmission used here. At maximum transmission, the detector pixel receiving the greatest flux recorded $2\times10^{7}$\,ph/pixel/s, equivalent to $2.6 \times 10^{8}$\,keV/pixel/s. These measurements end well below the maximum flux the detector can accurately measure, $3.2 \times 10^{9}$ keV/pixel/s. This is well beyond the capabilities of photon counting detectors.

\subsection{Radiographs}

\begin{figure}[t!]
\centering
\subfloat[][]{
	\centering
	\includegraphics[width=0.3\textwidth]{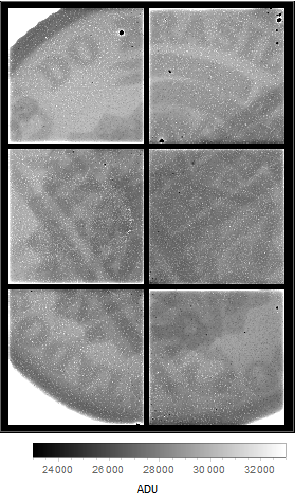}
}
\subfloat[][]{
	\centering
	\includegraphics[width=0.3\textwidth]{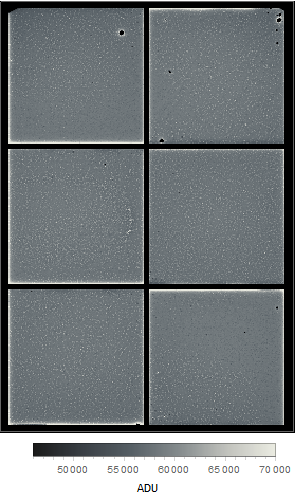}	
}
\subfloat[][]{
	\centering
	\includegraphics[width=0.3\textwidth]{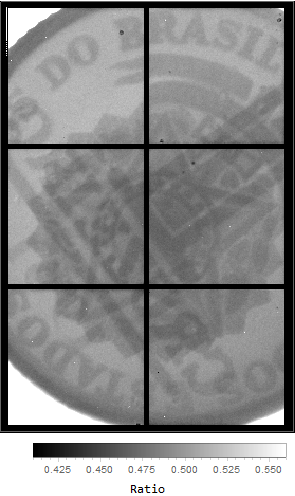}	
}
\caption{A low contrast image of an aluminum coin (Brazilian 1 Cruzeiro) taken with the CdTe MM-PAD and a silver anode microfocus x-ray tube with a bias voltage of 47\,kV, a filament current of 230\,$\upmu$A, and a 2\,mm aluminum x-ray filter. The displayed image is an average of one hundred 300\,ms images ($1.6 \times 10^{5}$\,x-rays/pixel total impinging on the coin in the dataset). The coin was placed close to the x-ray source to magnify the image ($\sim$ 3$\times)$. \textbf{(a)} The raw image from the detector with a simple dark frame subtraction. \textbf{(b)} A flat field image obtained by imaging the without the coin present. \textbf{(c)} The flat field corrected image, (image (a) divided by image (b)).}
    \label{fig:coin}
\end{figure}

Figure~\ref{fig:coin} shows a low contrast radiograph of an aluminum coin that was placed in close proximity to a microfocus silver anode x-ray tube biased to 47\,kV. The detector was placed to allow the geometrically magnified image of the coin to fill the detector ($\sim$ 3$\times$ magnification). Figure ~\ref{fig:coin}a is the raw radiograph. Figure~\ref{fig:coin}b shows an image taken with the same x-ray source without the coin. This flood illumination shows the non-uniformity inherent to CdTe sensor material, as well as characteristic dark defects (displayed as dark pixels in the lower tiles of the detector). Figure~\ref{fig:coin}c shows the ratio of a and b. This flat field correction significantly improves uniformity. The flat field correction is generally a function of x-ray energy because it depends on the average depth of x-ray conversion in the sensor material and lateral fields and/or traps that distort image. This can be an important consideration for polychromatic x-ray sources because samples can harden the transmitted beam significantly.

\begin{figure}[t!]
    \centering
    \includegraphics[width=0.9\textwidth]{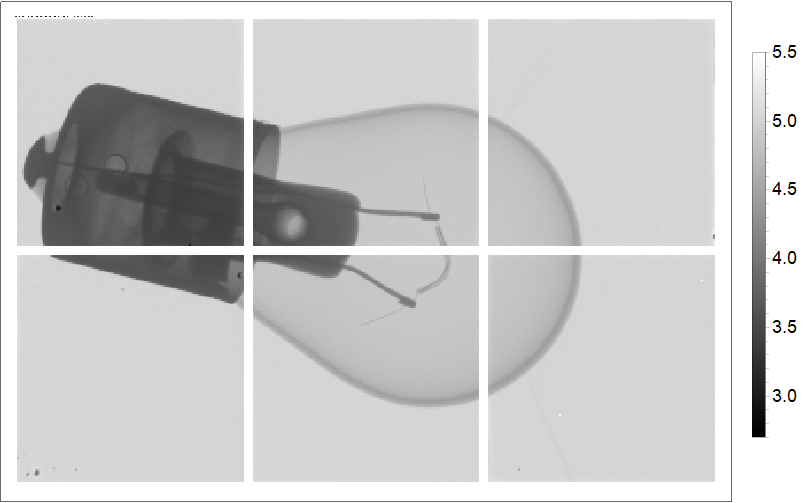}
    \caption{Log scale image of a light bulb with no flat field correction. The image was taken with a silver anode microfocus x-ray tube biased to 47\,keV, with an electron current of 500\,$\upmu$A, and 1\,mm aluminum x-ray filter. The displayed image is an average of one hundred 800 ms images. A total of $3.2 \times 10^{5}$\,x-rays/pixel illuminated the frame over the data set. In the most absorbing portions, < 2\% of the x-rays are transmitted.}
    \label{fig:bulb}
\end{figure}

Figure~\ref{fig:bulb} is a significantly higher contrast image of a light bulb as is shown without flat field correction. 
Nonuniformity in the flood illumination images is due to several effects. There are some sensitivity variations, but there are also significant lateral electric fields in the sensor material which cause lateral displacements in the recorded signal, and hence produce areal variations from pixel to pixel on the 1 percent level. For radiography, one properly corrects by dividing the flood illumination. However, since pixel areal variations are integrated intensity conserving, correcting diffraction data by simple division with a flood field may introduce some error~\cite{Green2013a}; see~\cite{Barna1999}\cite{Gruner2002} for a discussion of this effect and methods of calibration. In many cases the effect is small.
Quantitative analysis of the CdTe sensor response can be found in \cite{Becker2016b}.

%
%
%

\subsection{Polarization}

CdTe is well known to exhibit dose dependent polarization effects. We have observed and characterized several consequences due to polarization, namely the reduction of signal (count rate deficit) with dose, a deterioration of image uniformity with time and the lateral displacement of signal due to differential dosing~\cite{Becker2016a}. 
These effects begin to be noticeable when an incident x-ray fluence of $10^{11}$ keV/mm$^{2}$ is absorbed by the detector and become quite pronounced when the fluence is greater than $10^{12}$ keV/mm$^{2}$. Onset of polarization can be reduced by operating at a higher sensor bias. 400\,V was chosen for operation in this case as a compromise between polarization immunity and the number of pixels which exhibited fluctuating leakage current.

The polarization in the material can be reset through a change in bias applied to the sensor. Our procedure is as follows: Apply a slight forward bias (-2\,V for 60\,s), followed by a return to the operating voltage (400\,V in this case). We wait an additional 30\,s for the sensor to reach a steady state before taking new images.

\begin{figure}[b!]
    \centering
    \includegraphics[width=1\textwidth]{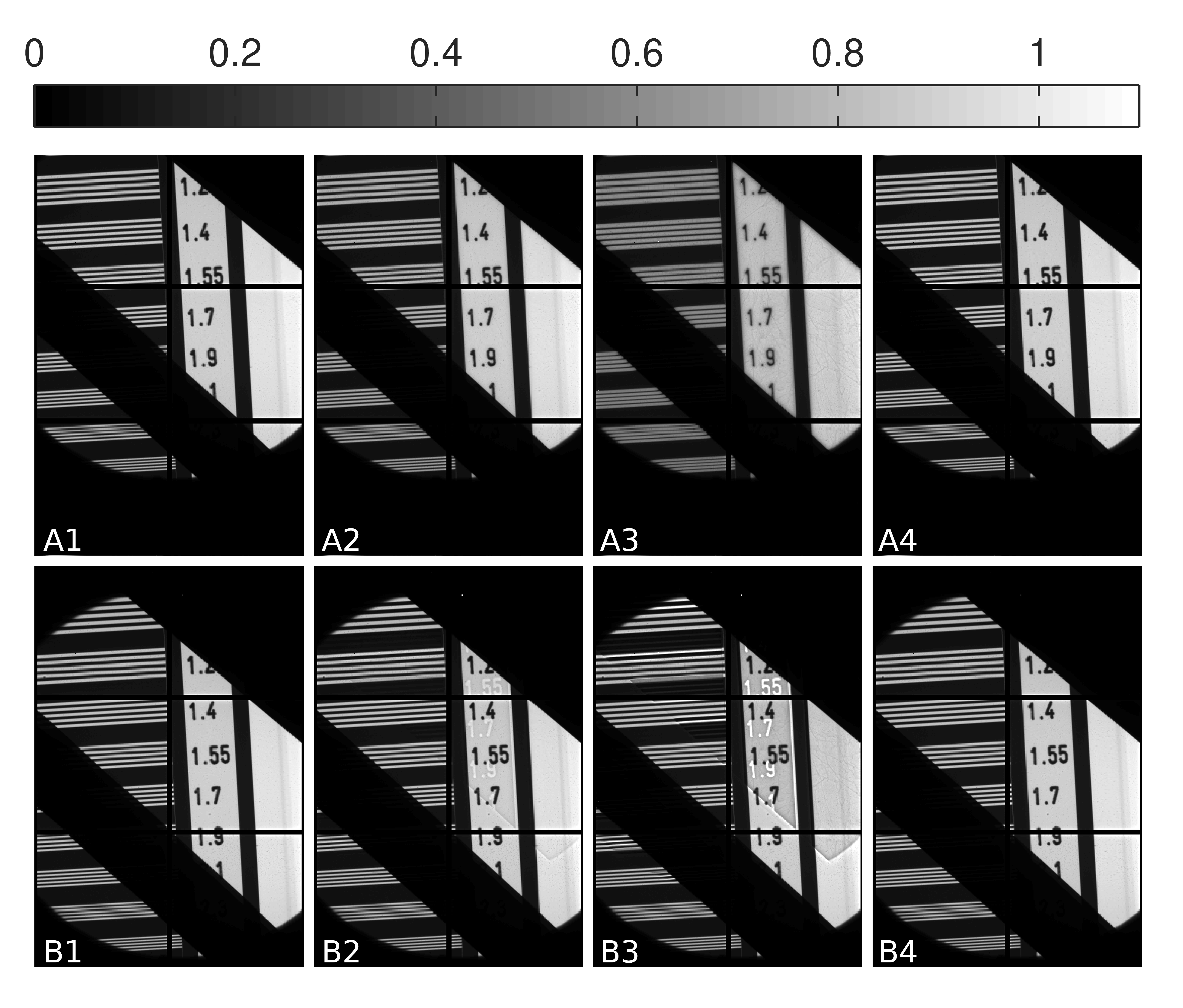}
    \caption{Onset of sensor polarization and efficacy of reset procedure. Images shown are of a standard x-ray line-pair mask taken at two different positions on the detector (A1-A4 at one position, B1-B4 displaced vertically from A by 4\,mm). Except for the brief time for exposing images at position B, all dose was accumulated at position A. Images taken after accumulated dose as follows: (A1,B1) - fresh detector; (A2,B2) - a dose of $1 \times 10^{10}$\,x-rays/mm$^{2}$; (A3,B3) - a dose of $8 \times 10^{10}$\,x-rays/mm$^{2}$; (A4,B4).
- after a new depolarization cycle. 
Panel A3 shows a loss in charge collection, a degredation to image resolution (due to lateral fields) and increased nonuniformity in the image. B3 highlights the reduction in charge collection in the dosed regions, but also shows lateral electric field effects through the increased charge collection of pixels at feature boundaries (notice ghosting of numbers which show a brightness greater than 1 in this panel). Images normalized to the signal in the brightest portions of panel A1.}
    \label{fig:polarization}
\end{figure}

Figure~\ref{fig:polarization} shows a sequence of images taken with a standard x-ray line-pair mask (cut from lead foil). These images were taken at two positions (displaced verically by 4\,mm) to highlight various features in the polarization progression. Except for a brief period to monitor at position B in the figure, all dose was accumulated at position A. Panel A3 shows a decrease in charge collection efficiency, a decrease in the uniformity of response, and a decrease in the resolution of the image. This last effect is due to the induced lateral fields which occur at the boundaries of exposed and unexposed regions. Panel B3 shows the resolution is maintained when a feature is moved into an uniformly polarized region, $e.g.,$ where the numbers imprinted on the mask have moved into a flat flood region. Notice also the ghosting at the old number positions. The intensity is higher than expected for a unpolarized region due to a lateral electric field pushing charge into these pixels from neighboring pixels. After a depolarization cycle, the response of the system returns to its initial state. 

Displacement of images with many high contrast features is a tool which especially highlights changes due to polarization. We note the polarization sensitivity is often more noticeable with a high-dynamic range imager, but is likely present at the same magnitude in any imager made with this material. We have chosen an aggressive de-polarization step to improve fidelity.

\section{Conclusions}

The CdTe MM-PAD detector has demonstrated exceptional capability as a high efficiency quantitative imager for high energy (>\,20\,keV) x-rays. The MM-PAD ASIC combines a charge integrating front-end with threshold driven charge dumps to achieve within each image both the single x-ray detection capability performance expected of photon counters~\cite{Giewekemeyer2014} and an instantaneous count-rate performance which exceeds that of photon counting imagers~\cite{Giewekemeyer2019}, all while framing at a kHz rate. These characteristics make the detector ideally suited for many difficult time resolved experiments, such as the temporal response of individual metal grains to periodic stress~\cite{Chatterjee2017,Chatterjee2019}, and ptychography, both in the sparse and high flux regimes~\cite{Giewekemeyer2014}\cite{Ayyer2014}.

To our knowledge, all available CdTe materials display non-ideal behavior that can lead to image distortion and charge collection inefficiency, phenomena collectively known as ``polarization''~\cite{Giewekemeyer2014}\cite{Ayyer2014}. Users should be aware that polarization effects occur in all detectors that use CdTe as the x-ray sensor material.  Depolarization by periodically cycling the voltage across the CdTe sensor mitigates these effects significantly and for many applications the magnitude of non-ideal performance does not limit scientific observation~\cite{Becker2016a}\cite{Becker2016b}. As an example, integration of bright x-ray Bragg spots can mask small scale distortions of details within the spot while yielding good integral values. Care should always be taken to limit the total integrated dose on all regions of the sensor between depolarization cycles.

In sum, the CdTe MM-PAD described in this paper is a valuable, productive tool in the arsenal of scientific x-ray imagers.

\section*{Acknowledgments}

X-ray detector development at Cornell is supported by U.S. Department
of Energy grant DE-SC0017631 and the Cornell High Energy
Synchrotron Source (CHESS) under U.S. National Science Foundation
award DMR-1332208. The authors appreciate the assistance of Zahir Islam and Russell Woods of Argonne National Lab's Advanced Photon Source while collecting high flux data used to characterize the CdTe MM-PAD. This research used resources of the Advanced Photon Source, a U.S. Department of Energy (DOE) Office of Science User Facility operated for the DOE Office of Science by Argonne National Laboratory under Contract No. DE-AC02-06CH11357.

%
%
%
%
%

\bibliographystyle{unsrt}
\bibliography{MasterBibFile}

\begin{thebibliography}{10}

\bibitem{Tate2013}
M~W Tate, D~Chamberlain, K~S Green, H~T Philipp, P~Purohit, C~Strohman, and S~M
  Gruner.
\newblock A medium-format, mixed-mode pixel array detector for kilohertz x-ray
  imaging.
\newblock {\em Journal of Physics: Conference Series}, 425(6):062004, mar 2013.

\bibitem{Giewekemeyer2014}
Klaus Giewekemeyer, Hugh~T. Philipp, Robin~N. Wilke, Andrew Aquila, Markus
  Osterhoff, Mark~W. Tate, Katherine~S. Shanks, Alexey~V. Zozulya, Tim Salditt,
  Sol~M. Gruner, and Adrian~P. Mancuso.
\newblock High-dynamic-range coherent diffractive imaging: ptychography using
  the mixed-mode pixel array detector.
\newblock {\em Journal of Synchrotron Radiation}, 21(5):1167--1174, aug 2014.

\bibitem{Giewekemeyer2019}
K.~Giewekemeyer, A.~Aquila, N.-T.~D. Loh, Y.~Chushkin, K.~S. Shanks, J.T.
  Weiss, M.~W. Tate, H.~T. Philipp, S.~Stern, P.~Vagovic, M.~Mehrjoo, C.~Teo,
  M.~Barthelmess, F.~Zontone, C.~Chang, R.~C. Tiberio, A.~Sakdinawat, G.~J.
  Williams, S.~M. Gruner, and A.~P. Mancuso.
\newblock Experimental 3d coherent diffractive imaging from photon-sparse
  random projections.
\newblock {\em {IUCrJ}}, 6(3):357--365, mar 2019.

\bibitem{Ayyer2014}
Kartik Ayyer, Hugh~T Philipp, Mark~W Tate, Veit Elser, and Sol~M Gruner.
\newblock Real-space x-ray tomographic reconstruction of randomly oriented
  objects with sparse data frames.
\newblock {\em Optics express}, 22(3):2403--2413, 2014.

\bibitem{Liu2017}
JP~Liu, J~Kirchhoff, L~Zhou, M~Zhao, MD~Grapes, DS~Dale, MD~Tate, HT~Philipp,
  SM~Gruner, TP~Weihs, et~al.
\newblock X-ray reflectivity measurement of interdiffusion in metallic
  multilayers during rapid heating.
\newblock {\em Journal of synchrotron radiation}, 24(4), 2017.

\bibitem{Overdeep2018}
Kyle~R Overdeep, Howie Joress, Lan Zhou, Ken~JT Livi, Sara~C Barron, Michael~D
  Grapes, Katherine~S Shanks, Darren~S Dale, Mark~W Tate, Hugh~T Philipp,
  et~al.
\newblock Mechanisms of oxide growth during the combustion of al: Zr
  nanolaminate foils.
\newblock {\em Combustion and Flame}, 191:442--452, 2018.

\bibitem{Tate2016}
Mark~W Tate, Prafull Purohit, Darol Chamberlain, Kayla~X Nguyen, Robert Hovden,
  Celesta~S Chang, Pratiti Deb, Emrah Turgut, John~T Heron, Darrell~G Schlom,
  et~al.
\newblock High dynamic range pixel array detector for scanning transmission
  electron microscopy.
\newblock {\em Microscopy and Microanalysis}, 22(1):237--249, 2016.

\bibitem{Jiang2018}
Yi~Jiang, Zhen Chen, Yimo Han, Pratiti Deb, Hui Gao, Saien Xie, Prafull
  Purohit, Mark~W. Tate, Jiwoong Park, Sol~M. Gruner, Veit Elser, and David~A.
  Muller.
\newblock Electron ptychography of 2d materials to deep sub-{\aa}ngström
  resolution.
\newblock {\em Nature}, 559(7714):343--349, jul 2018.

\bibitem{Chatterjee2017}
K~Chatterjee, JYP Ko, JT~Weiss, HT~Philipp, J~Becker, P~Purohit, SM~Gruner, and
  AJ~Beaudoin.
\newblock Study of residual stresses in ti-7al using theory and experiments.
\newblock {\em Journal of the Mechanics and Physics of Solids}, 109:95--116,
  2017.

\bibitem{Chatterjee2019}
K.~Chatterjee, A.~J. Beaudoin, D.~C. Pagan, P.~A. Shade, H.~T. Philipp, M.~W.
  Tate, S.~M. Gruner, P.~Kenesei, and J.-S. Park.
\newblock Intermittent plasticity in individual grains: A study using high
  energy x-ray diffraction.
\newblock {\em Structural Dynamics}, 6(1):014501, jan 2019.

\bibitem{Limousin2003}
O.~Limousin.
\newblock New trends in {CdTe} and {CdZnTe} detectors for x- and gamma-ray
  applications.
\newblock {\em Nuclear Instruments and Methods in Physics Research Section A:
  Accelerators, Spectrometers, Detectors and Associated Equipment},
  504(1-3):24--37, may 2003.

\bibitem{Becker2016b}
Julian Becker, Mark~W Tate, Katherine~S Shanks, Hugh~T Philipp, Joel~T Weiss,
  Prafull Purohit, Darol Chamberlain, Jacob~PC Ruff, and Sol~M Gruner.
\newblock Characterization of {CdTe} sensors with schottky contacts coupled to
  charge-integrating pixel array detectors for x-ray science.
\newblock {\em Journal of Instrumentation}, 11(12):P12013, dec 2016.

\bibitem{Becker2017}
Julian Becker, Mark~W Tate, Katherine~S Shanks, Hugh~T Philipp, Joel~T Weiss,
  Prafull Purohit, Darol Chamberlain, and Sol~M Gruner.
\newblock Sub-microsecond x-ray imaging using hole-collecting schottky type
  cdte with charge-integrating pixel array detectors.
\newblock {\em Journal of Instrumentation}, 12(06):P06022, 2017.

\bibitem{Aamir2011}
R~Aamir, S~P Lansley, R~Zainon, M~Fiederle, A~Fauler, D~Greiffenberg, P~H
  Butler, and A~P~H Butler.
\newblock Pixel sensitivity variations in a {CdTe}-medipix2 detector using
  poly-energetic x-rays.
\newblock {\em Journal of Instrumentation}, 6(01):C01059--C01059, jan 2011.

\bibitem{SchuettePHD}
Dan Schuette.
\newblock {\em A mixed analog and digital pixel array detector for synchrotron
  x-ray imaging}.
\newblock PhD thesis, Cornell University, 2008.

\bibitem{Green2013a}
Katherine~S Green, Hugh~T Philipp, Mark~W Tate, Joel~T Weiss, and Sol~M Gruner.
\newblock Calibration and post-processing for photon-integrating pixel array
  detectors.
\newblock {\em Journal of Physics: Conference Series}, 425(6):062009, mar 2013.

\bibitem{Barna1999}
S.~L. Barna, M.~W. Tate, S.~M. Gruner, and E.~F. Eikenberry.
\newblock Calibration procedures for charge-coupled device x-ray detectors.
\newblock {\em Review of Scientific Instruments}, 70(7):2927--2934, jul 1999.

\bibitem{Gruner2002}
Sol~M. Gruner, Mark~W. Tate, and Eric~F. Eikenberry.
\newblock Charge-coupled device area x-ray detectors.
\newblock {\em Review of Scientific Instruments}, 73(8):2815--2842, aug 2002.

\bibitem{Becker2016a}
Julian Becker, Mark~W. Tate, Katherine~S. Shanks, Hugh~T. Philipp, Joel~T.
  Weiss, Prafull Purohit, Darol Chamberlain, and Sol~M. Gruner.
\newblock High-speed imaging at high x-ray energy: {CdTe} sensors coupled to
  charge-integrating pixel array detectors.
\newblock In {\em AIP Conference Proceedings}, volume 1741, page 040037. AIP
  Publishing, 2016.

\end{thebibliography}

\end{document}